\RequirePackage{snapshot}
\documentclass[conference]{IEEEtran}
\IEEEoverridecommandlockouts
\usepackage{cite}
\usepackage{amsmath,amssymb,amsfonts}
\usepackage{algorithmic}
\usepackage{graphicx}
\usepackage{subcaption}
\usepackage{textcomp}
\usepackage{xcolor}
\usepackage{hyperref}
\usepackage[colorinlistoftodos]{todonotes}

\def\BibTeX{{\rm B\kern-.05em{\sc i\kern-.025em b}\kern-.08em
    T\kern-.1667em\lower.7ex\hbox{E}\kern-.125emX}}
\begin{document}

\title{The Adobe Hidden Feature\\ and its Impact on Sensor Attribution
}

\author{\IEEEauthorblockN{Butora Jan}
\IEEEauthorblockA{\textit{Univ. Lille, CNRS, Centrale Lille,} \\
\textit{ UMR 9189 CRIStAL,}\\
F-59000 Lille, France \\
jan.butora@cnrs.fr}
\and
\IEEEauthorblockN{Bas Patrick}
\IEEEauthorblockA{\textit{Univ. Lille, CNRS, Centrale Lille,} \\
\textit{ UMR 9189 CRIStAL,}\\
F-59000 Lille, France \\
patrick.bas@cnrs.fr}
}

\maketitle

\begin{abstract}
If the extraction of sensor fingerprints represents nowadays an important forensic tool for sensor attribution, it has been shown recently in~\cite{baracchi2021facing,albisani2021checking,iuliani2021leak} that images coming from several sensors were more prone to generate False Positives (FP) by presenting a common "leak". In this paper, we investigate the possible cause of this leak and after inspecting the EXIF metadata of the sources causing FP, we found out that they were related to the Adobe Lightroom or Photoshop softwares. The cross-correlation between residuals on images presenting FP reveals periodic peaks showing the presence of a periodic pattern. By developing our own images with Adobe Lightroom we are able to show that all developments from raw images (or 16 bits per channel coded) to 8 bits-coded images also embed a periodic $128\times128$ pattern very similar to a watermark. However, we also show that the watermark depends on both the content and the architecture used to develop the image. The rest of the paper presents two different ways of removing this watermark, one by removing it from the image noise component, and the other by removing it in the pixel domain. We show that for a camera presenting FP in \cite{iuliani2021leak}, we were able to prevent the False Positives. A discussion with Adobe representatives informed us that the company decided to add this pattern in order to induce dithering. 
\end{abstract}


\section{Motivations}
The use of the Photo-Response Non-Uniformity noise (PRNU) for imaging sensor attribution is one operational success coming from the forensic research with the seminal paper and associated patent of Lukas, Goljan and Fridrich in 2005~\cite{lukas2005determining,fridrich2010method}. It relies on the fact that each photo-site of the sensor is corrupted by a multiplicative noise (i.e. proportional to the noiseless value) which is the same for all acquisitions from the same sensor but completely different from one sensor to another. This noise survives the development processes and the mathematical model in the pixel domain can be written as:
\begin{equation}\label{eq:image}
    \mathbf{I} = \mathbf{I}^o+\mathbf{K}\mathbf{I}^o + \mathbf{\Theta},
\end{equation}
where $\mathbf{K}$, $\mathbf{I}^o$, $\mathbf{I}$, and $\mathbf{\Theta}$ denote respectively the PRNU component, the noiseless image, the captured image, and a collection of independent random noise components.

Estimating the PRNU component $\hat{\mathbf{K}}$ related to one given sensor enables to extract a fingerprint of this sensor which can be used in forensic tasks, to associate images captured with this sensor, or to detect manipulations on specific areas.

\begin{figure}[h]
    \begin{centering}
    \includegraphics[width=0.99\columnwidth]{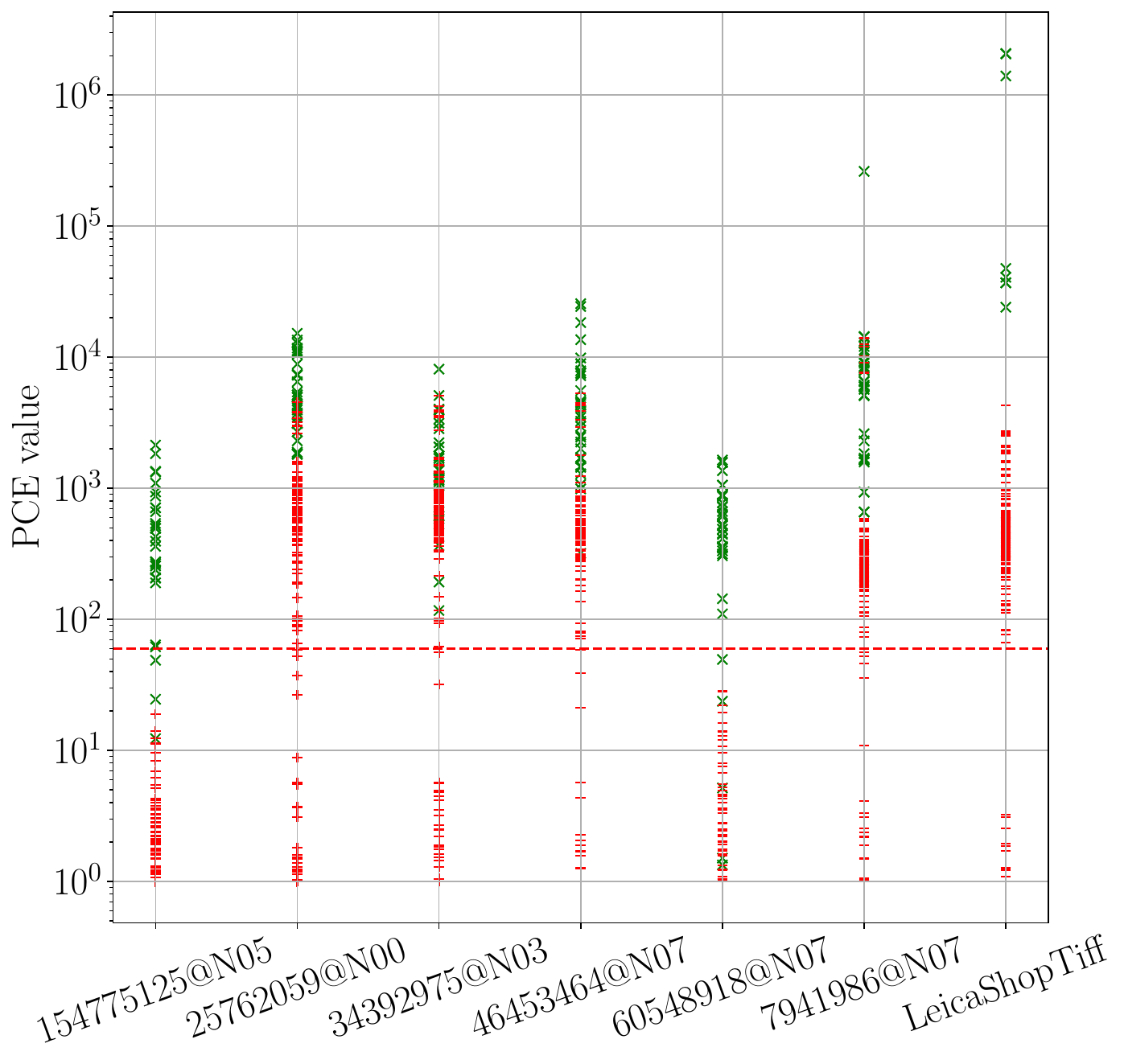}
    \end{centering}
    \caption{The original $\mathrm{PCE}$s for the Leica Q2 camera and 7 different devices. The first 6 labels are the FlickR identification names and the last one is images coming from a Q2 camera shared by the Leica Shop in Lille and developed using Adobe Lightroom in 8-bit Tiff format. Matching and mismatching tests are reported in green and red, respectively. The threshold of 60 is highlighted by the red dashed line.}
    \label{fig:Q2origPCE}
\end{figure}

For camera sensor attribution, the classical methodology to attribute a picture to a given sensor was benchmarked by Goljan {\it et al.}~\cite{goljan2009large} in 2009 on a database of $10^6$ JPEG images.

For a set of $N$ reference images $\{\mathbf{I}_1,\dots,\mathbf{I}_N \}$ sometimes called flat-field images (the estimation is better if these images are both out of focus and bright), the fingerprint $\hat{\mathbf{K}}$ is estimated using a maximum likelihood estimator~\cite{fridrich2009digital} applied on every image residual $\mathbf{W}_i$:
\begin{equation}\label{eq:fingerprint}
    \hat{\mathbf{K}}=\frac{\sum\limits_i \mathbf{W}_{i}\mathbf{I}_{i}}{\sum\limits_i {\mathbf{I}_{i}}^2},
\end{equation}
where the residual is computed as $\mathbf{W}_{i}=\mathbf{I}_{i}-f(\mathbf{I}_{i})$, $f(.)$ being a denoising function such as the ones proposed by Mihcak {\it et al.}~\cite{mihcak1999low} or Cherchia {\it et al.}~\cite{chierchia2014bayesian} which relies on the BM3D~\cite{dabov2009bm3d} denoising algorithm combined with Markov Random Field model.

In order to potentially attribute a test image $\mathbf{I}^t$ with the fingerprint $\hat{\mathbf{K}}$, the normalized correlation between the image residual $\mathbf{W}^t=\mathbf{I}^t-f(\mathbf{I}^t)$ and the potential fingerprint specific to the image $\hat{\mathbf{K}}\mathbf{I}^t$ is first computed:
\begin{equation}
\mathrm{NCC}(s_1,s_2)=\frac{<\mathbf{W}^t(s_1,s_2);\hat{\mathbf{K}}\mathbf{I}^t>}{|\mathbf{W}^t(s_1,s_2)|.|\hat{\mathbf{K}}\mathbf{I}^t|},
\end{equation}
where $(s_1,s_2)$ represents the spatial shift, which could arise due to cropping. Eventually, the statistic which is used to decide whether or not the image can be attributed to the fingerprint $\hat{\mathbf{K}}$ is the Peak to Correlation Energy ($\mathrm{PCE}$) defined as:
\begin{equation}\label{eq:PCE}
    \mathrm{PCE}=\frac{\mathrm{NCC}(s_1^\mathrm{peak},s_2^\mathrm{peak})^2}{\frac{1}{mn-|\mathcal{N}|}\sum\limits_{(s_1,s_2) \notin \mathcal{N}} \mathrm{NCC}(s_1,s_2)^2},
\end{equation}
where $\mathcal{N}$ denotes a small neighborhood centered on the maximum of the cross-correlation function located at $(s_1^\mathrm{peak},s_2^\mathrm{peak})$, and $(m,n)$ are the dimensions of the correlation function. On a large scale database presented in~\cite{goljan2009large}, the authors proposed an attribution threshold w.r.t. the $\mathrm{PCE}$ of 60, which in the setup of the reference paper is associated with a practical FP rate of $2.4\times 10^{-5}$ without considering potential translations on the test image. While the neighborhood size was proposed to $11\times11$ region, we use only $2\times2$ neighborhood.\\

Recently different papers studied this attribution procedure on modern sensors coming either from recent digital cameras or smartphones, and they found out that the benchmark proposed in 2009~\cite{goljan2009large} was now subject to numerous FPs. A major overview of this problem was proposed by Iuliani {\it et al.}~\cite{iuliani2021leak} by considering 33K pictures uploaded on the FlickR photo-sharing platform coming from 45 smartphones and 25 modern digital cameras. This study exhibited important FP rates (i.e. $>5\%$) for smartphones such as the {\it iPhone 11 pro}, the {\it Huawei P20 pro} or {\it Mate 20 Pro}, the {\it Samsung Galaxy A50}, the {\it Nokia Pureview 808}, or the {\it Xiaomi Redmi Note 7}; and for digital cameras such as the {\it Canon M6 Mark II}, the {\it Fuji X-T30}, the {\it Leica Q2}, the {\it Nikon D780} or {\it Z50} or {\it the Sony DSC-RX0}.\\

Complementary works partially analyzed the causes of these wrong attributions and ways to anticipate potential false positives.
In~\cite{albisani2021checking} Albisani {\it et al.} show that some FP were associated with smartphone captures in {\it portrait} mode, i.e. presenting an out-of-focus background generated artificially.
In~\cite{goljan2009large} Baracchi {\it et al.} focused on captures in portrait mode coming from the {\it iPhone X} and proposed a way to mitigate the wrong estimation of the fingerprint in the background by weighting the fingerprint w.r.t. the depth map associated with the capture.
In~\cite{bhat2022investigating}, Bhat and Bianchi show that steganalysis features such as SPAM~\cite{pevny2010steganalysis} can be used to detect smartphones presenting potential biases added during the image development pipeline.
In~\cite{liu2023fits} Liu {\it et al.} consider a specific noise coming from the software and they propose to mitigate it by decreasing the $\mathrm{PCE}$ values by a constant $C$ specific to the camera when these values under the null hypothesis present a strong bias. Note that this solution requires an {\it a priori} knowledge of the bias $C$.

\begin{figure}[t]
    \begin{centering}
    \includegraphics[width=0.99\columnwidth]{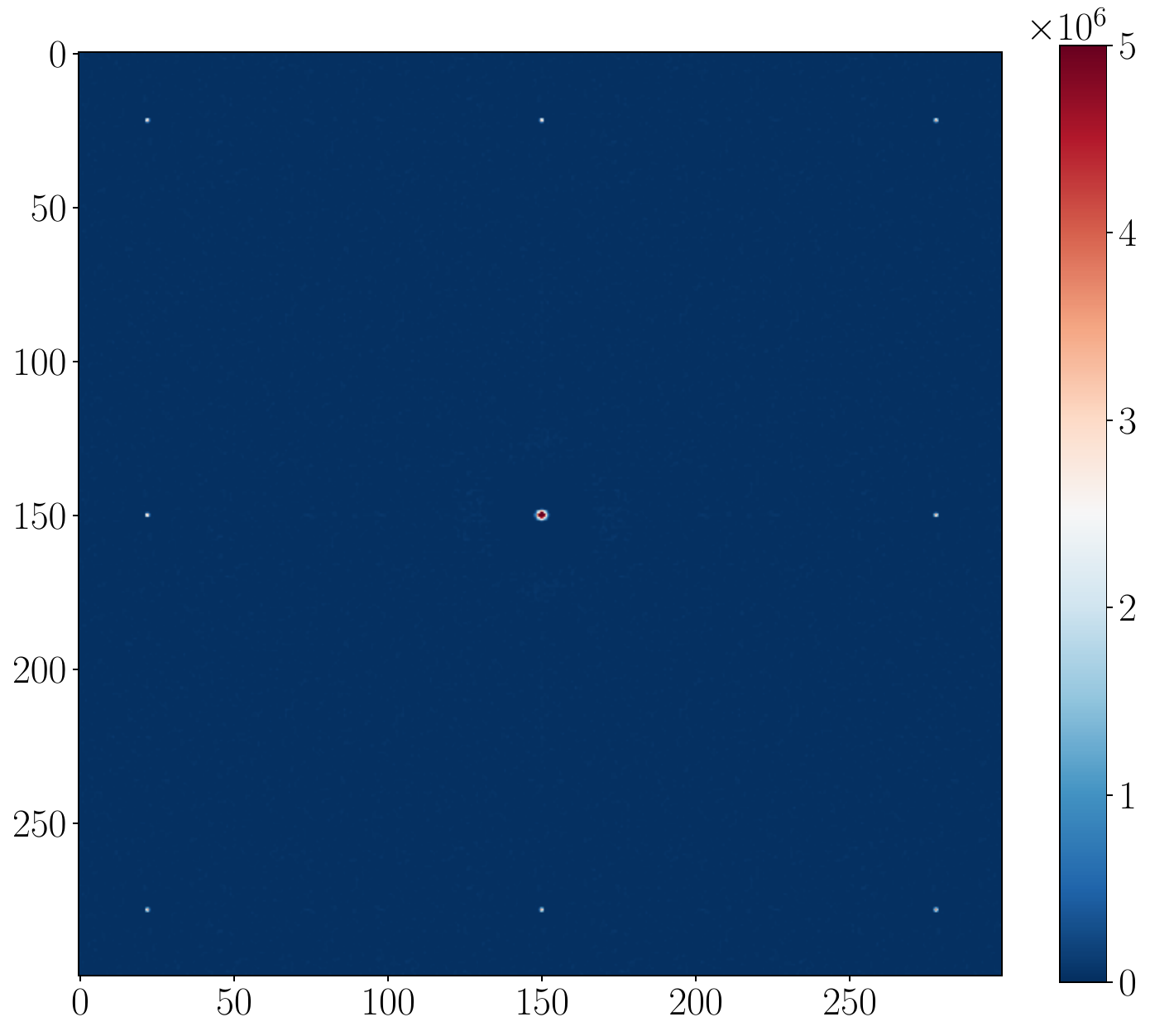}
    \end{centering}
    \caption{Crop of the autocorrelation function of one residual associated with a source generating FP.}
    \label{fig:crossCorrFing}
\end{figure}

\section{The process discovery}

This section (and the beginning of the next one) is presented as a story, which means that the style is not very formal or academic. It reflects how the authors experienced this research, going from assumptions to surprising discoveries.

Before starting this analysis, we had several ideas in mind regarding the possible causes of FP using the PRNU.
Since the fingerprint $\mathbf{K}\mathbf{I}^o$ is always added to the noiseless image, it can be seen as a "bias" on the original image. False positives consequently needed to be associated with extra biases, which are often named Non Unique Artefacts (NUAs) by the forensics community. The two main origins of a bias we could imagine were: 1) the JPEG dimples~\cite{dimples20} found by Agarwal and Farid {\it et al.} which are due to hardware implementations of quantization strategies during the JPEG compression, and 2) adding a constant noise during the capture or in the image development pipeline.

We decided to focus on the Leica Q2 camera and we wanted first to confirm the results presented in~\cite{iuliani2021leak}.

Using the python PRNU implementation from Bondi {\it et al.}~\cite{PRNUPython} but taking care of computing the $\mathrm{NCC}$ w.r.t. $\hat{\mathbf{K}}\mathbf{I}^t$ and not only $\hat{\mathbf{K}}$ as in the original implementation, we were able to reproduce the results as depicted in Fig.~\ref{fig:Q2origPCE}.

What was surprising was the fact that out of the seven different sources, only five of them were subject to FP. After a deeper inspection of the EXIF metadata using verbose outputs\footnote{this can be achieved using the command \texttt{exiftool -v5 image.jpg}} of the different sources, we found out that the sources generating FP were not Out Of Camera (OOC) JPEGs but all had a tag related to Adobe software, either Adobe Lightroom or Adobe Photoshop.

\begin{figure}[t]
    \begin{centering}
        \includegraphics[width=1\columnwidth]{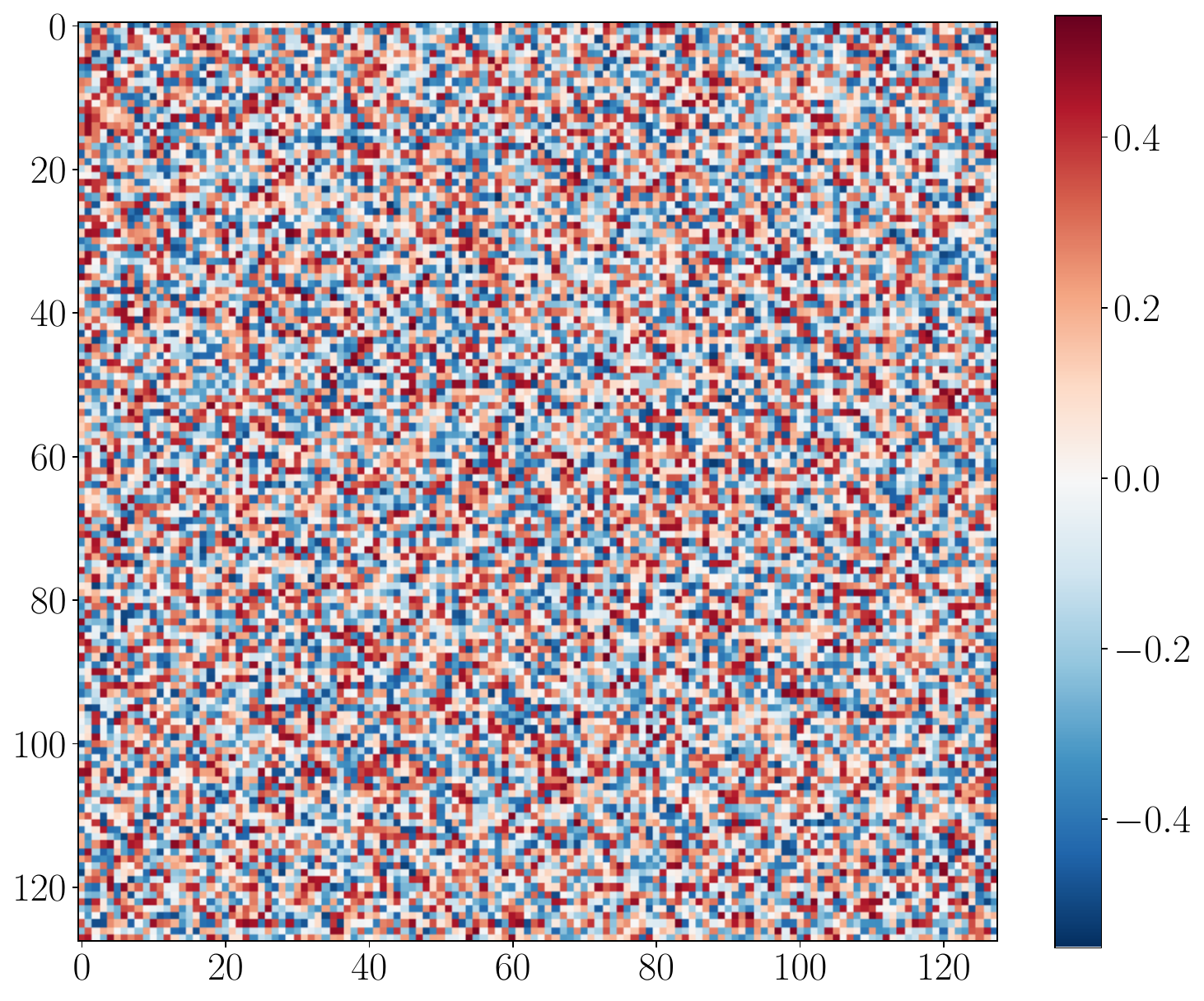}
    \par\end{centering}
    \caption{\label{fig:tif_residual}
    Average of non-overlapping $128\times 128$ patches of an image residual.
    }
\end{figure}

We then looked at the autocorrelation function of one residual belonging to one of these sources and as illustrated in Fig.~\ref{fig:crossCorrFing} we were able to clearly see periodic peaks on a $128\times 128$ grid. Finally averaging $128\times 128$ patches taken from the very same grid exhibited very similar patterns on the different sources coming from Adobe (an example of the average pattern is depicted in Fig.~\ref{fig:tif_residual}).

The final "spit it out" test was to develop using either Adobe Lightroom or Photoshop one constant image saved in a RAW format (DNG) or a 16-bit tiff. For both formats, a periodic $128\times 128$ pattern was present on the produced jpeg or 8-bit image except on Adobe Photoshop when we directly exported in 8 bits per channel PNG format.
This was noticeable for all the different OS we tested (iOS, macOS, Windows). Other tests confirm the fact that the pattern is added on each RGB component independently and that is is independent of the image content.

Once we were convinced that the signal which can be considered as a {\it watermark} was embedded by Adobe Lightroom or Photoshop, we performed different tests to understand at which step of the development process was the watermark embedded. We noticed that the watermark is not dependent on the processes that the image could undergo (e.g. rotation, sharpening, denoising, ...), which means that the watermark was not added in the photo-site domain but on the contrary just before the conversion from 16 bits per channel to 8 bits per channel.

Last but not least, the watermark was not present on exported tiff images in 16 bits per channel.

\begin{figure}[h]
    \begin{centering}
        \begin{subfigure}{0.49\columnwidth}
            \includegraphics[width=0.95\columnwidth]{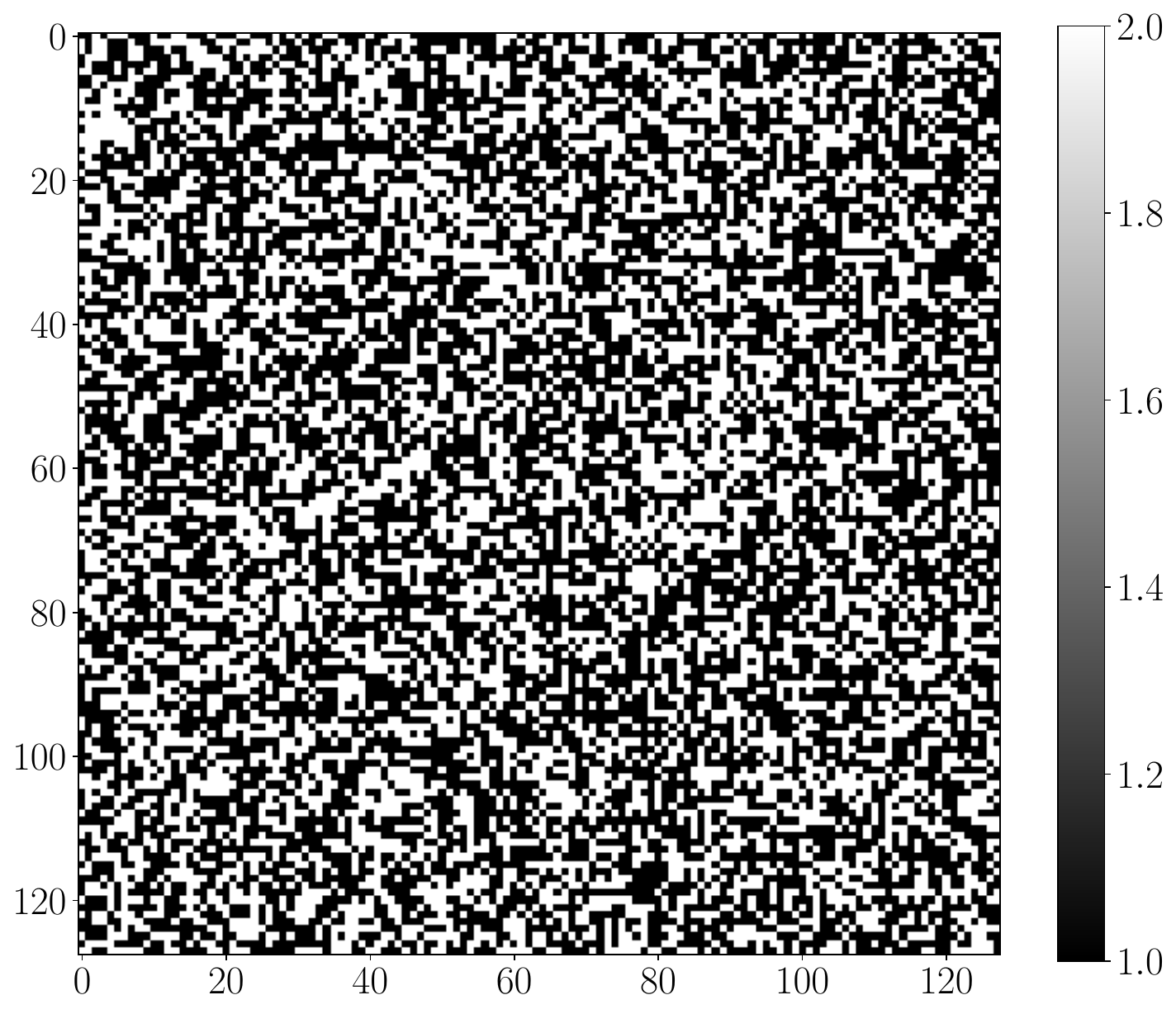}
            \caption{val = 382}
        \end{subfigure}
        \begin{subfigure}{0.49\columnwidth}
            \includegraphics[width=0.95\columnwidth]{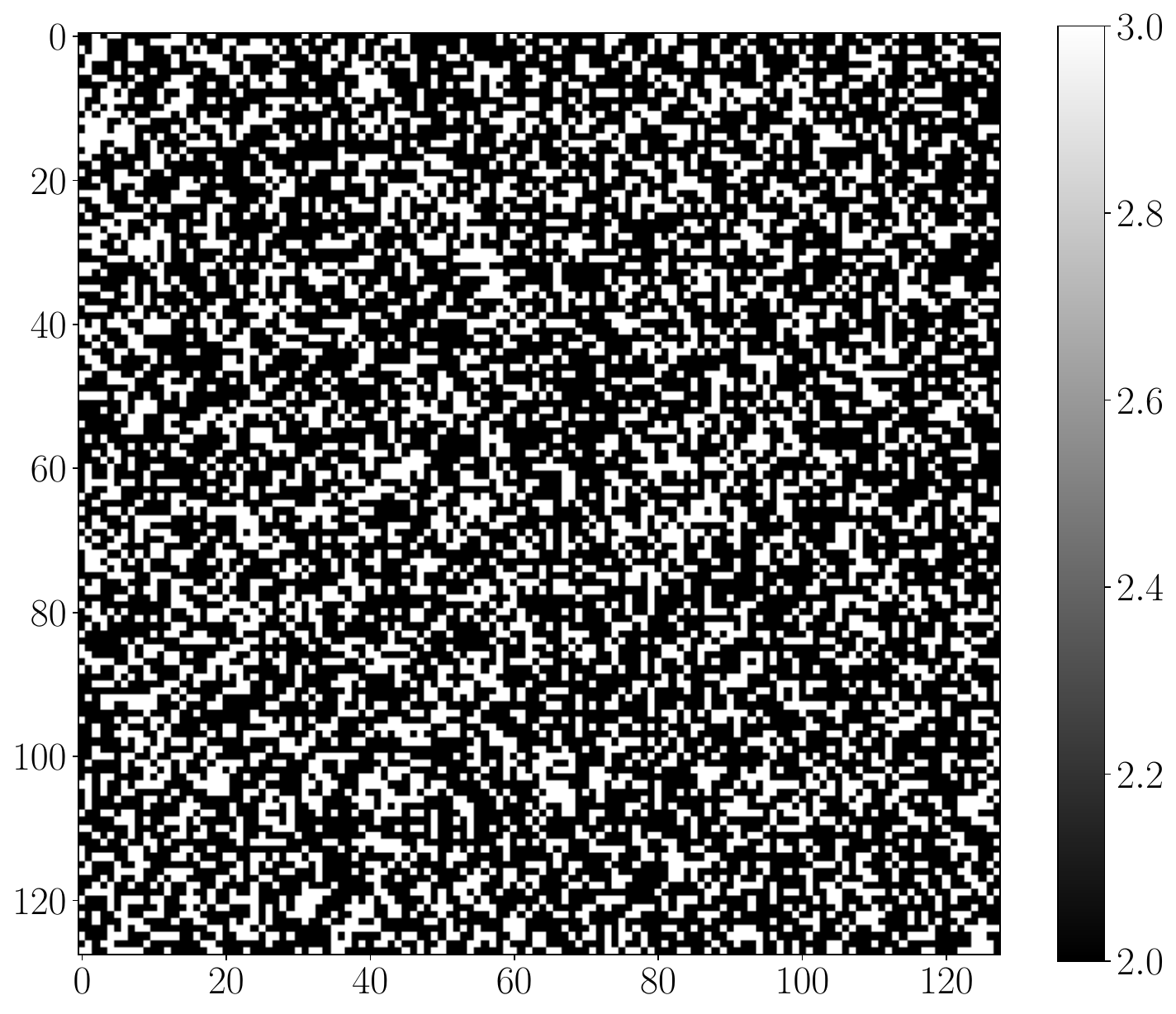}
            \caption{val = 638}
        \end{subfigure}
        \begin{subfigure}{0.49\columnwidth}
            \includegraphics[width=0.95\columnwidth]{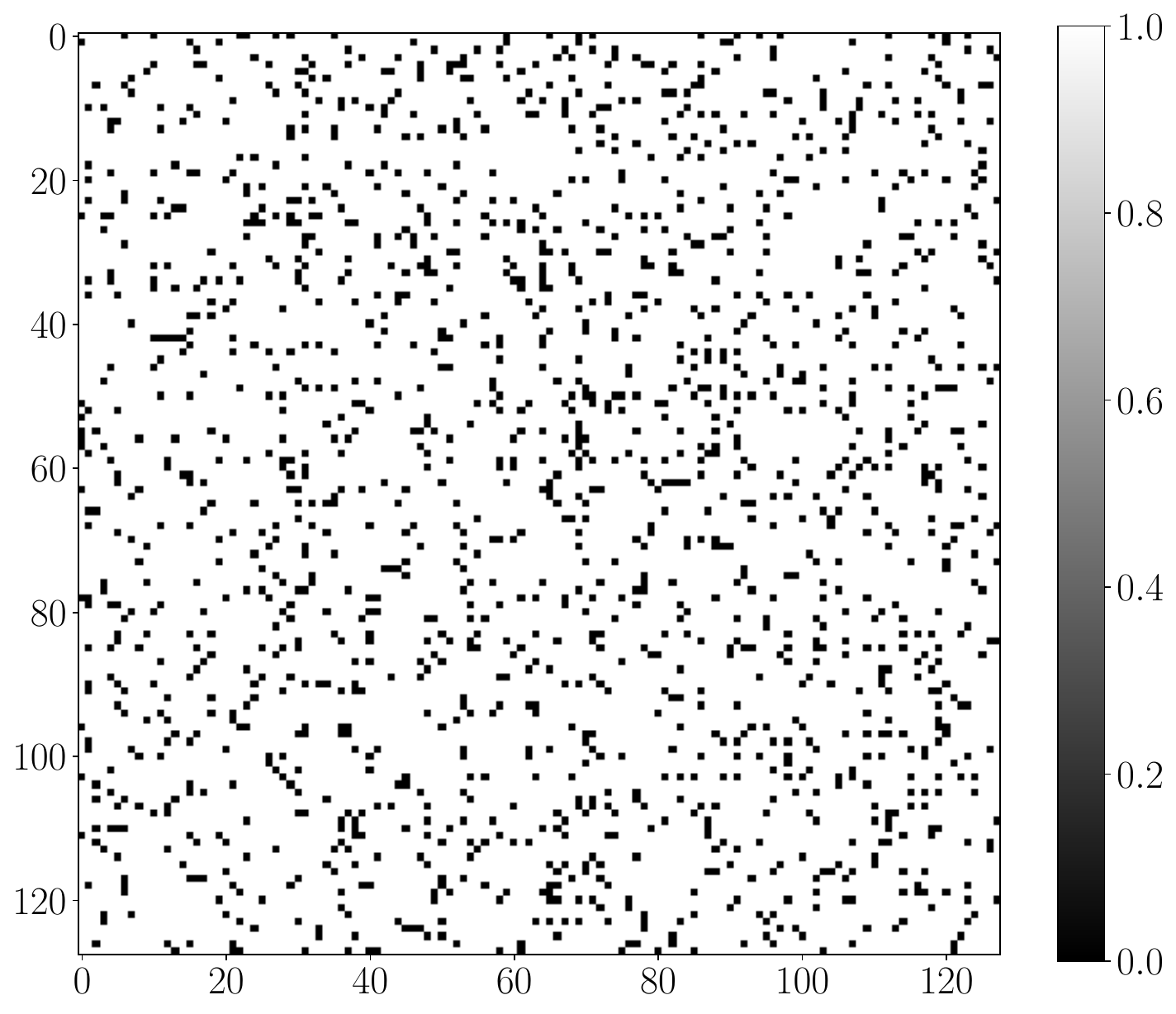}
            \caption{Difference}
        \end{subfigure}
    \par\end{centering}
    \caption{\label{fig:const_images}
    Changes introduced after 16bit $\rightarrow$ 8bit quantization of two constant TIFF images of size $128\times 128$. (a): 382 (16bit) $\rightarrow$ 1 (8bit), (b): 638 (16bit) $\rightarrow$ 2 (8bit). (c): Their difference image.}

    \label{fig:16-8quant}
\end{figure}

\begin{figure}[t]
    \begin{centering}
        \begin{subfigure}{0.7\columnwidth}
            \includegraphics[width=1\columnwidth]{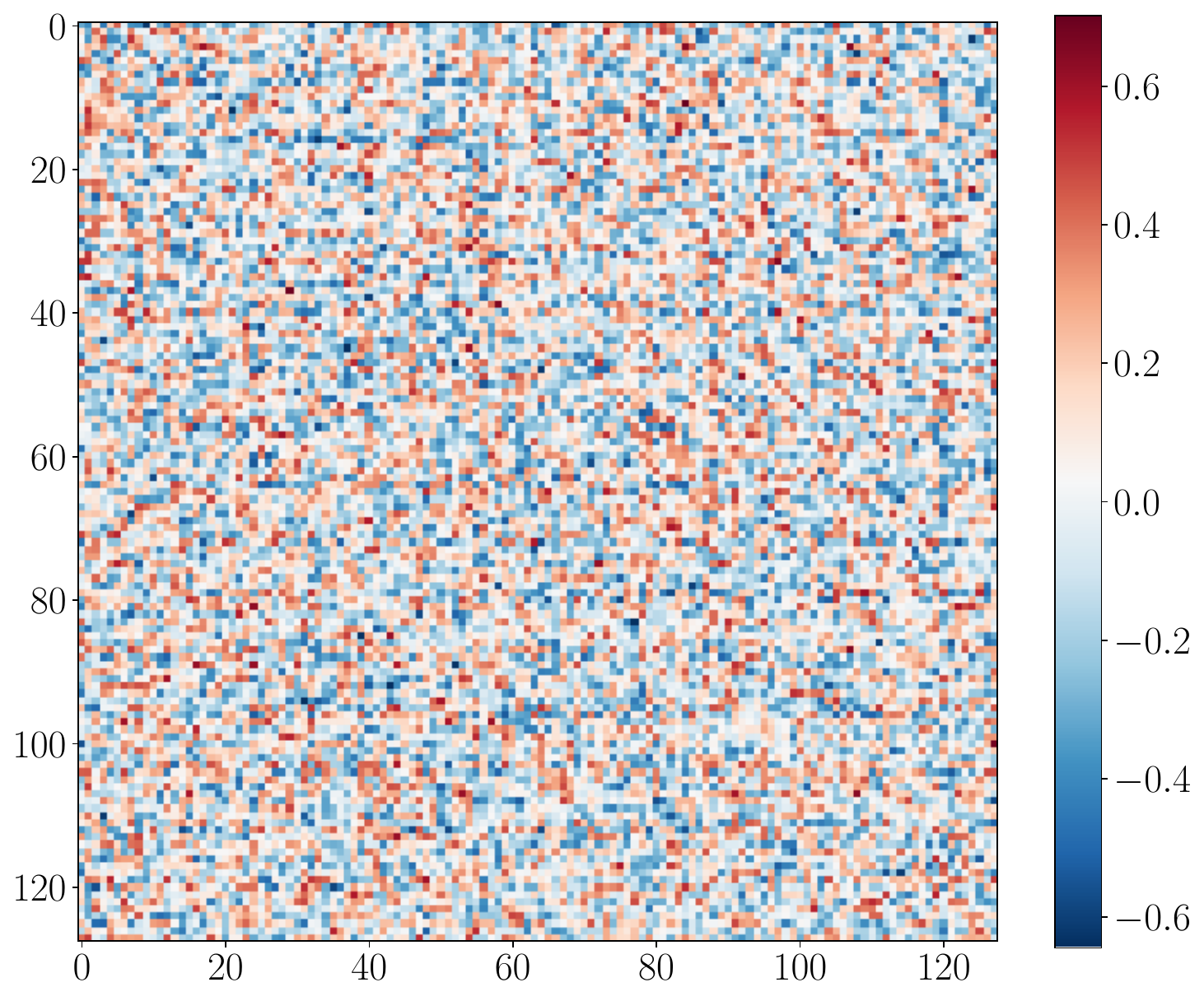}
            \caption{Watermark estimated on M1 chip. (QF100)}
            \label{subfig:M1}
        \end{subfigure}
        \begin{subfigure}{0.7\columnwidth}
            \includegraphics[width=1\columnwidth]{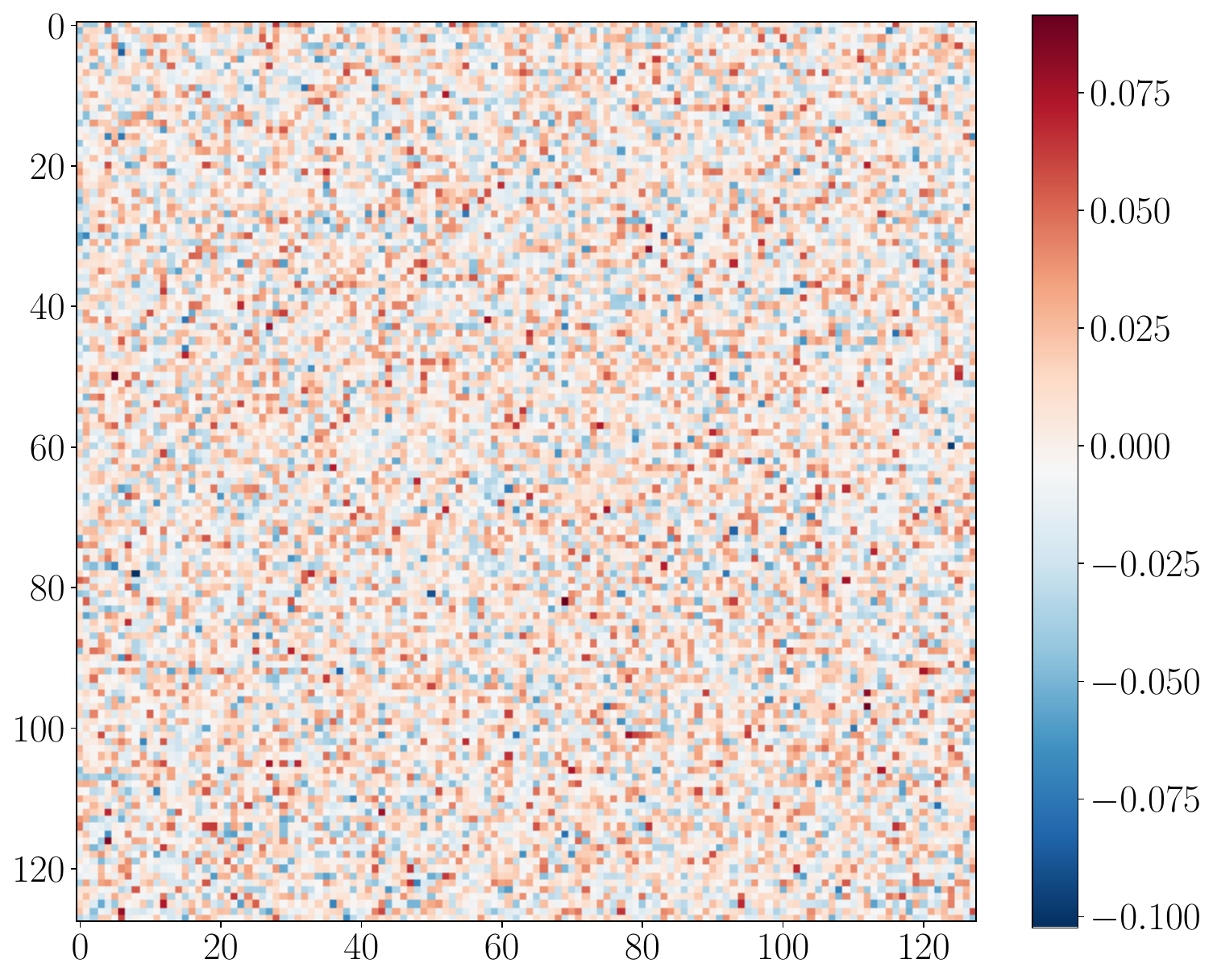}
            \caption{Difference w.r.t. an Intel chip (QF 100).}
            \label{subfig:diffIntel}
        \end{subfigure}
        \begin{subfigure}{0.7\columnwidth}
            \includegraphics[width=1\columnwidth]{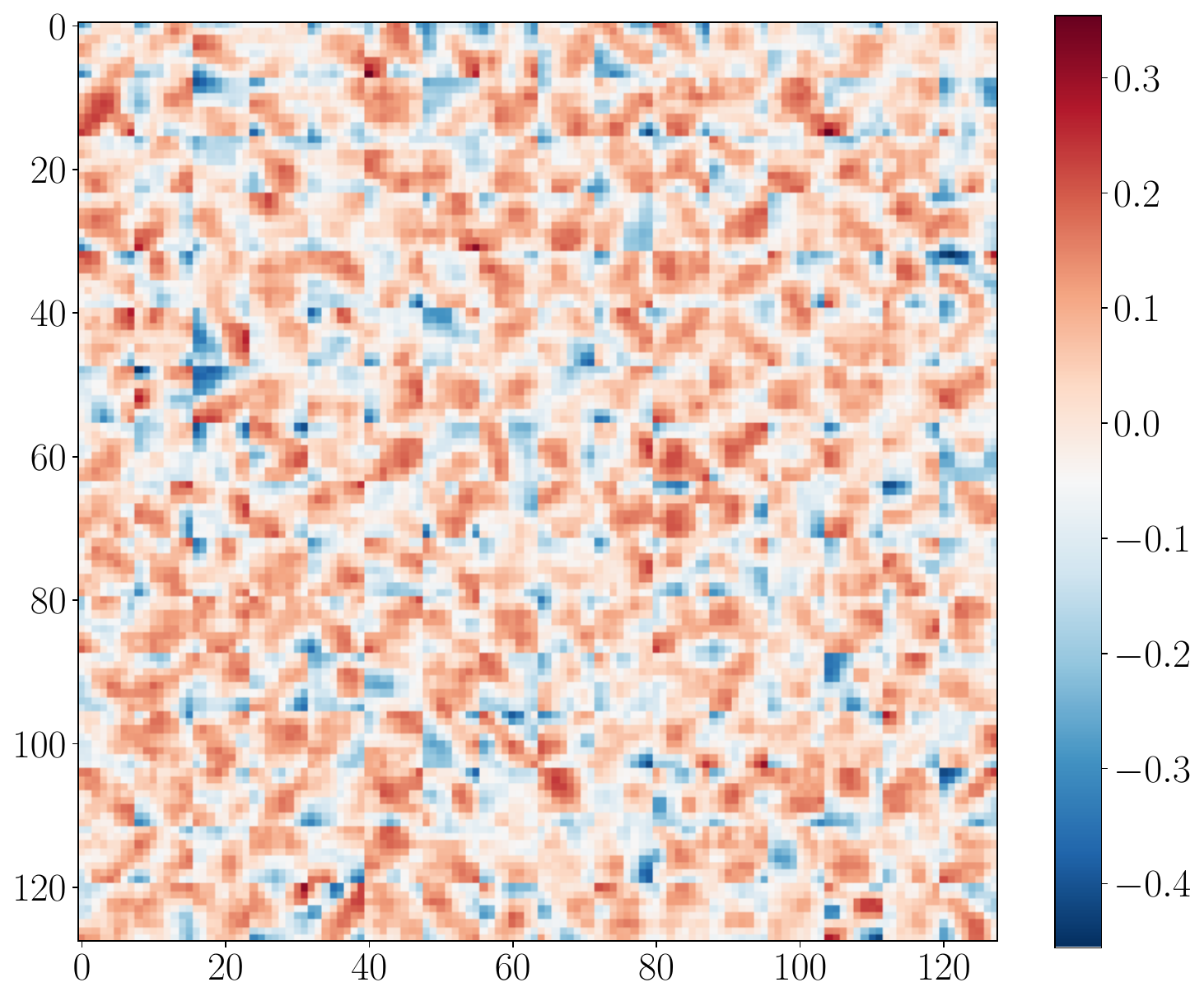}
            \caption{Watermark estimated at QF 80.}
            \label{subfig:M1_80}
        \end{subfigure}
    \par\end{centering}
    \caption{\label{fig:dif_architectures}
    For a constant RAW image with added Gaussian white noise with $\sigma=2$, developed with Lightroom to JPEG, we show (a) the estimated watermark when developed to QF 100 on Apple M1 chip, (b) the difference between estimated watermarks at QF 100 when the same image was developed with Apple M1 chip and Intel chip, (c) estimated watermark when developed to QF 100 on Apple M1 chip.}
\end{figure}



\section{Watermark Properties}\label{sec:properties}

Once we were sure that the embedding of a watermark was responsible for the FPs obtained in the PRNU attribution, our goal was to find a way to remove it in order to prevent FP. From a watermarking security perspective~\cite{bas2016watermarking}, the security scenario seems similar to a Constant Message Attack (CMA)~\cite{cayre2005watermarking} (the same watermark is present in all the watermarked documents), and in this case, the attack is straightforward: once the watermark has been properly estimated, a simple subtraction of this signal enables to remove it.

Unfortunately, our first tests revealed that it is not possible to remove the watermark by a plain subtraction of one unique $128\times128$ periodical pattern on the image, even when considering the impact of JPEG compression.  This is due to at least three factors:
\begin{enumerate}
    \item The fact that the watermark is embedded in the 16-bit domain, where it is constant, but then suffers quantization to 8 bits. Depending on the value of the pixel component in the 16-bit domain, the modifications in the 8-bit domain are different. As illustrated in Fig.~\ref{fig:16-8quant} for constant images and for two different values in the 16-bit domain, the quantized watermarks are correlated but slightly different, even though the values $382,638$ are equal modulo $256$.
    \item The fact that the image development pipeline is different between one architecture and another. This feature was not expected and prevented us from estimating one average unique watermark (as done for the sensor fingerprint using~\eqref{eq:fingerprint}) and removing it. Depending on the CPU/GPU used, but also on the OS, we noticed that the watermark changed by about 10\%. Fig.~\ref{subfig:M1} shows one estimated average $128\times 128$ watermark estimated on an Apple computer with a M1 chip. and Fig.~\ref{sub@subfig:diffIntel} the difference w.r.t. the watermark generated from an Intel chip. In both cases, the number of $128\times 128$ patches is such that the estimation error is negligible. Note that to estimate an expectation of the watermark after 16 to 8-bit quantization, we add a small Gaussian noise on the RAW image.
    \item The fact that the watermark is also shaped by the JPEG coder which depends on parameters such as the quality factor. Fig.~\ref{subfig:M1} and Fig.\ref{subfig:M1_80} show the estimation of the watermark for two different quality factors used in Adobe Lightroom (which are different from the standard ones). Here again, the two watermarks are considerably different.
\end{enumerate}

While it is possible in principle to reverse-engineer the watermark for 8-bit tiff images and a given architecture (one has to develop $2^{16}$ constant 16-bit tiff images of size $128\times128$), it becomes computationally unfeasible for JPEG images due to dependencies within a block of 64 pixels and different quantization tables.
To sum up, in order to correctly remove the watermark, we reached the conclusion that we need to estimate the watermark for each source (i.e. each set of template images used to estimate the camera fingerprint). We show in the next section two strategies to remove it and consequently prevent the occurrence of FP during sensor attribution.






\section{Removing the Watermark}
In this section, we introduce two methods for removing the watermark from the estimated PRNUs.

\begin{figure}[h]
    \begin{centering}
    \includegraphics[width=0.6\columnwidth]{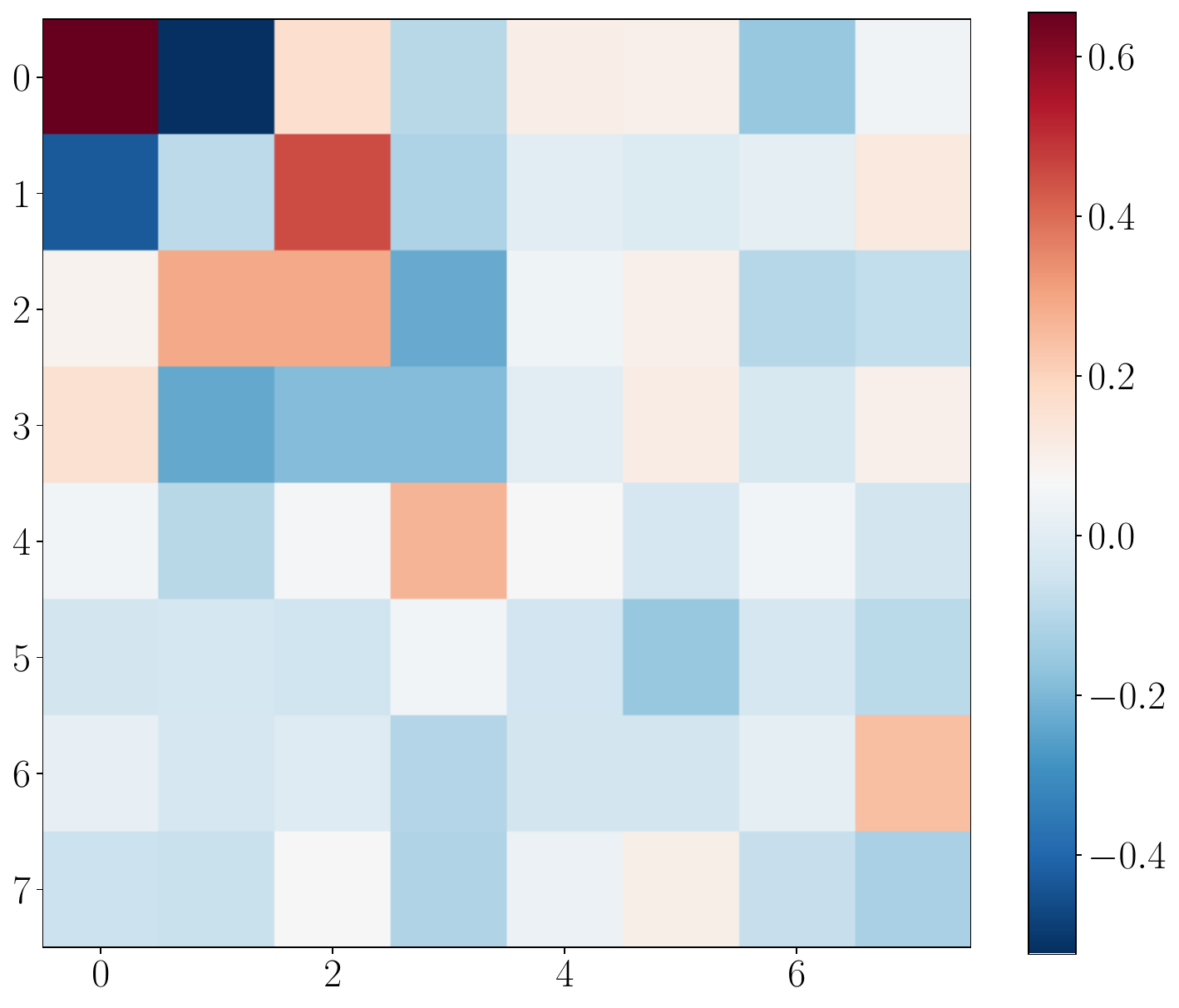}
    \par\end{centering}
    \caption{\label{fig:dimples}Estimate of JPEG dimples in the pixel domain from a single image taken with the Q2 camera JPEG compressed with QF 100.}
\end{figure}

Before we start to explain how to remove the watermark generated by Adobe, we mention another class of Non Unique Artefacts (NUAs) represented by the JPEG dimples~\cite{dimples20} and how to estimate and remove them. In short, dimples can create a non-zero bias on non-overlapping $8\times8$ patches, which also can be seen as a periodic pattern. See Fig.~\ref{fig:dimples} for the bias introduced by the dimples in the pixel domain estimated from a single image taken with the Q2 camera.

To efficiently remove the watermark $\mathbf{w}_\mathbf{I}\in\mathbb{R}^{128\times128}$,\footnote{We put the image $\mathbf{I}$ into subscript to emphasize the watermark's dependency on the image content.} we introduce its effect, together with the JPEG dimples $\mathbf{d}\in \mathbb{R}^{8\times8}$ into the model of the developed image~\eqref{eq:image}. To ease the notation, we assume in the following that all variables are grayscale patches of size $128\times128$, which can be achieved by taking non-overlapping crops of the images (or copying the $8\times8$ dimples several times) and by converting potential color images into grayscale. Our model of the watermarked image is then:
\begin{equation}\label{eq:watermarked_image}
    \mathbf{I} = \mathbf{I}^o+\mathbf{K}\mathbf{I}^o + \mathbf{d} + \mathbf{w}_\mathbf{I} + \mathbf{\Theta}.
\end{equation}

\subsection{Removing the dimples}

Before removing the watermark, we remove the JPEG dimples that can be see nas additional periodic pattern. We estimate the dimples by averaging all non-overlapping $8\times 8$ patches $\mathbf{I}_p\in\mathcal{P}_8$ of the image:

\begin{equation}\label{eq:dimple}
    \hat{\mathbf{d}} = \frac{1}{|\mathcal{P}_8|}\sum_{\mathbf{I}_p\in\mathcal{P}_8} \mathbf{I}_p.
\end{equation}

We then remove the bias created by the dimples by simply subtracting the estimate~\eqref{eq:dimple} from every image patch, $\mathbf{I}'_p = \mathbf{I}_p - \hat{\mathbf{d}}$. We will refer to the dimple-free image as $\mathbf{I}'$.

\subsection{Canceling the component in the image residual}

With the watermarked model of the image~\eqref{eq:watermarked_image}, we can now express the image residual as:

\begin{align}
    \mathbf{W} & = \mathbf{I}' - f(\mathbf{I}') \nonumber\\
    & = \mathbf{I}\hat{\mathbf{K}} + \hat{\mathbf{w}_\mathbf{I}} + \mathbf{\Theta}  \nonumber\\
    & \quad + (\mathbf{I}^o-f(\mathbf{I}')) + (\mathbf{I}^o-\mathbf{I})\mathbf{K} + (\mathbf{w}_\mathbf{I} - \hat{\mathbf{w}_\mathbf{I}}) + (\mathbf{d} - \hat{\mathbf{d}}) \nonumber\\
    & = \mathbf{I}\hat{\mathbf{K}} + \hat{\mathbf{w}_\mathbf{I}} + \mathbf{\Omega}, \label{eq:watermarked_residual}
\end{align}

where~$\hat{\mathbf{w}_\mathbf{I}}$ is the estimate of the watermark, and~$\mathbf{\Omega}$ is a sum of the five accumulated independent noise components.

Next, we estimate the expected watermark from the (dimple-free) residual $\mathbf{W}$, as the average $128\times 128$ patch:

\begin{equation}\label{eq:watermark}
\hat{\mathbf{w}} = \frac{1}{|\mathcal{P}_{128}|}\sum_{\mathbf{w}_p\in\mathcal{P}_{128}} \mathbf{w}_p.
\end{equation}
Note that this is only the average watermark, and every patch $p$ carries a possibly different realization of this watermark due to its dependency on the image content, as mentioned in Section~\ref{sec:properties}. We thus employ a modified version of the Gram-Schmidt orthogonalization process in order to remove a potentially different watermark from every patch. Let $\mathbf{W}'$ the dimple-free residual patch orthogonalized against the expected watermark $\hat{\mathbf{w}}$. A Gram-Schmidt orthogonalization states that
\begin{equation} \label{eq:GS_process}
    \mathbf{W}' = \mathbf{W} - proj_{\hat{\mathbf{w}}}(\mathbf{W})\hat{\mathbf{w}},
\end{equation}

where $proj_{\hat{\mathbf{w}}}(\mathbf{W})=\frac{<\mathbf{W}, \hat{\mathbf{w}}>}{\|\hat{\mathbf{w}}\|^2}$. However, in a case where $proj_{\hat{\mathbf{w}}}(\mathbf{W})<0$, the update~\eqref{eq:GS_process} adds a positive multiple of the expected watermark, which is undesirable. To this end, we modify the projection to
\begin{equation}\label{eq:projection}
    proj_{\hat{\mathbf{w}}}(\mathbf{W}) = \max \left(0, \frac{<\mathbf{W}, \hat{\mathbf{w}}>}{\|\hat{\mathbf{w}}\|^2}\right),
\end{equation}

thus the final estimate of the watermark can be expressed as
\begin{equation}\label{eq:watermark_estimate}
    \hat{\mathbf{w}}_\mathbf{I}=proj_{\hat{\mathbf{w}}}(\mathbf{W})\hat{\mathbf{w}}.
\end{equation}

Finally, we want to point out that if the dimples or the watermark are not present, the estimate $\hat{\mathbf{w}_\mathbf{I}},\hat{\mathbf{d}}$ will be very close to zero.

Following the same reasoning as in~\cite{lukas2005determining}, observing $\mathbf{W}, \mathbf{I}, \hat{\mathbf{w}}_\mathbf{I}, \hat{\mathbf{d}}$, the ML estimate $\hat{\mathbf{K}}$ for the PRNU can be found as
\begin{equation}\label{eq:PRNU_residual}
    \hat{\mathbf{K}}=\frac{\sum\limits_i \left(\mathbf{W}_{i}  - \hat{\mathbf{w}}_{\mathbf{I}_i}\right)\mathbf{I}_{i}}{\sum\limits_i {\mathbf{I}_{i}}^2},
\end{equation}

where the subtractions are performed on every $8\times8$ and $128\times 128$ patch respectively.

\begin{figure}[h]
    \begin{centering}
    \includegraphics[width=\columnwidth]{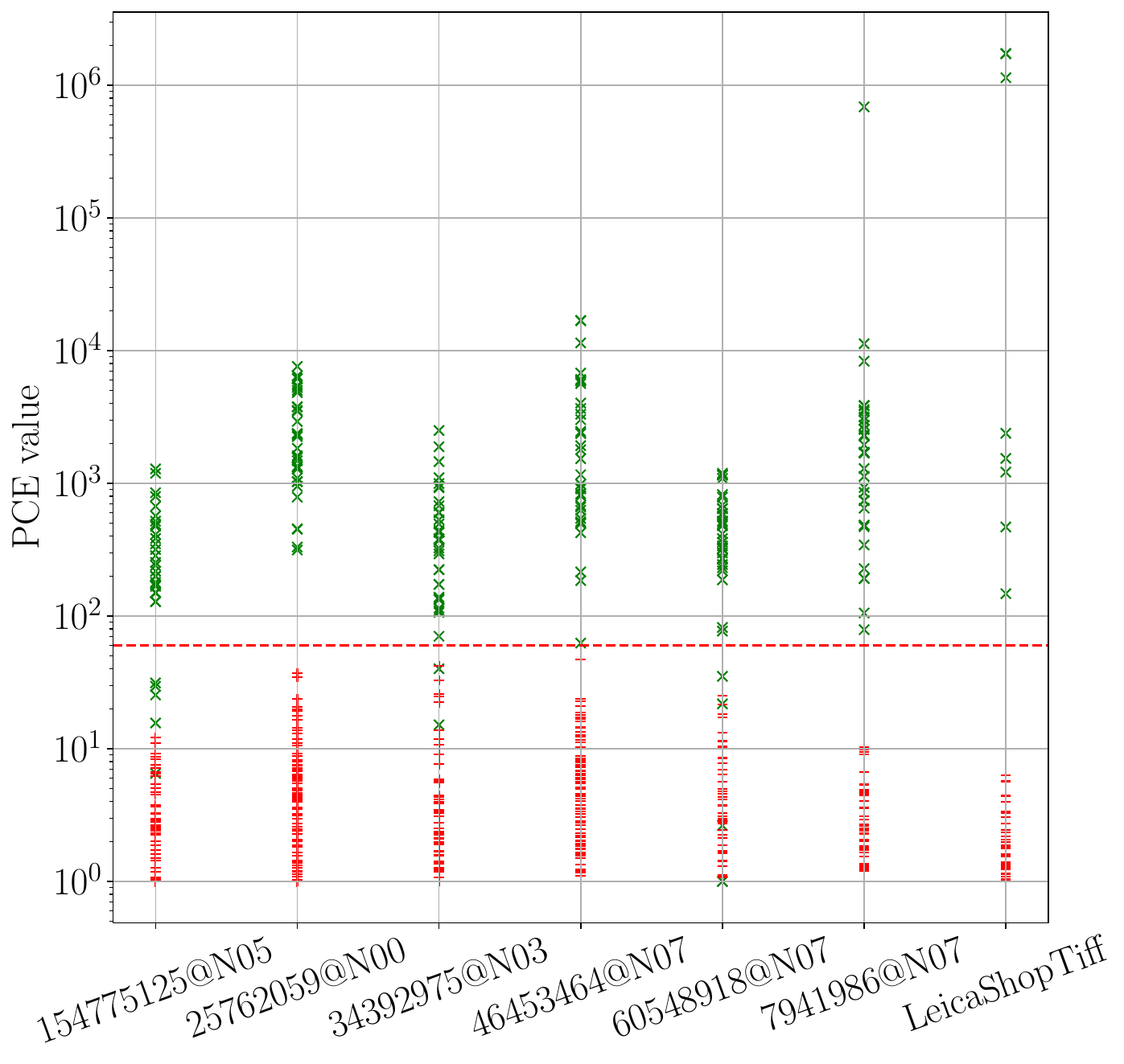}
    \par\end{centering}
    \caption{\label{fig:Q2_PCE_residual}
    PCE statistics computed from Leica Q2 camera with the proposed \emph{residual} canceling. Matching and mismatching tests are reported in green and red, respectively. The threshold of 60 is highlighted by the red dashed line.}
\end{figure}

The results with the PRNU~\eqref{eq:PRNU_residual} are shown in Fig.~\ref{fig:Q2_PCE_residual} and we can observe that the proposed methodology effectively mitigates all the False Positives previously present (see Fig.~\ref{fig:Q2origPCE}). The code used to generate the results is available at \url{https://github.com/janbutora/prnu-python}.

\subsection{Cancelling the component in the image}
An alternative approach is to remove the watermark from the observed image $\mathbf{I}$ itself, which could be more desirable for forensics users wanting to continue to process the image.

For simplicity, we use the estimates of the average watermark $\hat{\mathbf{w}}$ and dimples $\hat{\mathbf{d}}$ from the previous Section. However, the estimate of the watermark in a given patch is computed from the dimple-free image $\mathbf{I}'$ as:
\begin{equation}
    \hat{\mathbf{w}}_{\mathbf{I}'} = proj_{\hat{\mathbf{w}}}(\mathbf{I}')\hat{\mathbf{w}},
\end{equation}

where the $proj$ function is defined in~\eqref{eq:projection}. The dimple-free, watermark-free image is then
\begin{equation}\label{eq:clean_image}
    \mathbf{I}''=\mathbf{I}'-\hat{\mathbf{w}}_{\mathbf{I}'}.
\end{equation}

\begin{figure}[h]
    \begin{centering}
    \includegraphics[width=\columnwidth]{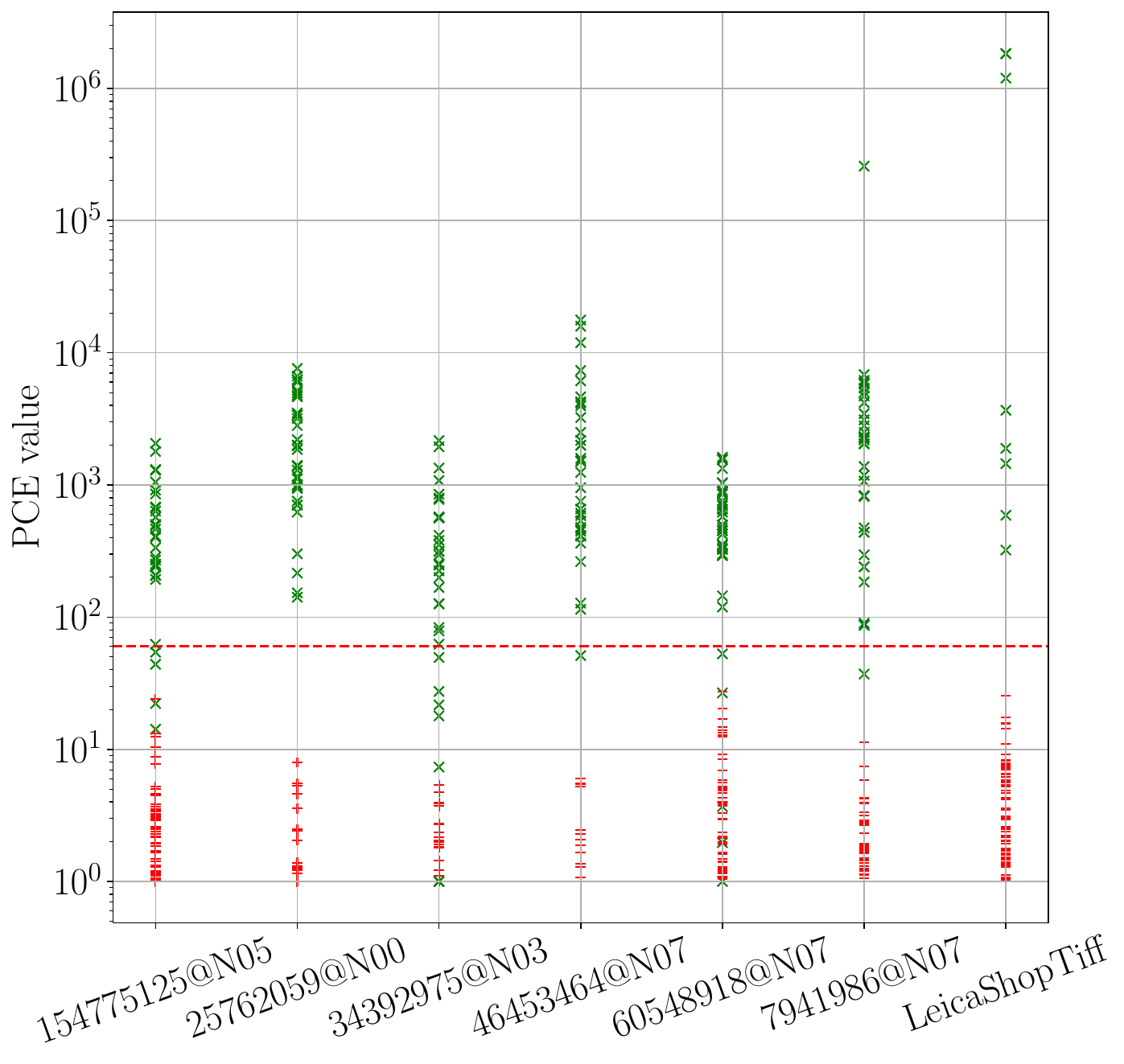}
    \par\end{centering}
    \caption{\label{fig:Q2_PCE_spatial}
    PCE statistics computed from Leica Q2 camera with the proposed \emph{spatial} canceling. Matching and mismatching tests are reported in green and red, respectively. The threshold of 60 is highlighted by the red dashed line.}
\end{figure}

The ML estimate of the PRNU can be obtained as:
\begin{equation}\label{eq:PRNU_spatial}
    \hat{\mathbf{K}}=\frac{\sum\limits_i \mathbf{W}''_{i} \mathbf{I}''_{i}}{\sum\limits_i \left(\mathbf{I}''_{i}\right)^2},
\end{equation}

where $\mathbf{W}''_i=\mathbf{I}''_i-f(\mathbf{I}_i'')$ is the residual of the updated image.
The results with the PRNU~\eqref{eq:PRNU_spatial} are shown in Fig.~\ref{fig:Q2_PCE_spatial} and as previously, we can observe that although having slightly different results, the proposed method effectively prevents the False Positives.

We want to point out, that after the residual or the image has been stripped of the potential watermark and dimples using~\eqref{eq:GS_process} or~\eqref{eq:clean_image}, we do not need to update the test image or its residuals in order to compute the PCE value~\eqref{eq:PCE}, because the watermark/dimples of test images can be simply considered as independent noise components w.r.t. the sensor fingerprint. Consequently, we did not update the residuals in our experiments.



\section{Discussion}
In this paper, we accidentally reverse-engineered a part of Adobe Lightroom and found out that a pattern, very similar to a public watermark, was embedded by the software before the conversion from 16 bits to 8 bits. We know also that this watermark has been used since at least 2014\footnote{it was detected for example in this image: \href{https://www.flickr.com/photos/isaacyuphotography/15961122615}{https://www.flickr.com/photos/isaacyuphotography/15961122615} .}, and it appears in all the six different development engines proposed by Adobe Lightroom. We also have checked that for two different Adobe Lightroom users, the watermark remains identical. Note also that the watermark is also embedded by Adobe Camera Raw, the raw conversion tool of Adobe Photoshop, but not when exporting directly to 8 bits PNG images.

Because this embedding has an important impact on camera sensor attribution using PRNU, specifically by creating false positives, we decided to propose methods to remove the watermark and consequently help the forensic community.

As a scientific remark, it is interesting to notice that even if the additive embedding process is extremely simple, the fact that it is done in the 16-bit domain and that the watermark is furthermore processed by specific hardware and/or JPEG parameters makes its removal rather sophisticated. This is due to the fact that once processed by the development pipeline, it is no longer a constant pattern in the 8-bit domain and it needs to be specifically estimated before canceling it.

Now one question remains, why does Adobe add a $128 \times 128$ periodical pattern on each channel before conversion to 8 bits?

We have asked to people working on Adobe Camera Raw and Lightroom and they told us that the pattern is used as a way to perform dithering and prevent undesirable effects such as banding. They also informed us that this dithering function is implemented in the Adobe DNG SDK which can be used to develop RAW image in the DNG format\footnote{the DNG format stands for Digital NeGative and has been developed by Adobe and it a very popular raw format used by several camera manufacturers, but also iOS or Android devices.}, specifically within the \texttt{dng\_utils.cpp} function. 

Nevertheless, it is worth mentioning that this pattern can also be used to perform forensic analyses by locally detecting the presence of this pattern. Note however that the watermark embedding process is not secure since the watermark can easily be estimated as in any Constant Message Attack scenario~\cite{bas2016watermarking}, but its periodicity makes it very robust to classical geometrical transforms such as rotations, scaling, and cropping operations as the method proposed by Kutter in 1998~\cite{Kutter:1998:WRT}. 

As a closing remark, we have also noticed that within the FlickR database presented in~\cite{goljan2009large}, some digital cameras such as the {\it Nikon D780} or {\it Z50} generated FP which are not related to the Adobe Lightroom watermark because we have not observed any periodic patterns. However, another (non-periodic) watermark can still be present even in these cases. A more extensive analysis will consequently be needed to fully solve these problems.
\newpage

\section{Acknowledgements}
The authors would like to thank Patrick De Smet (NICC, ENFSI DIWG) for bringing the subject of false positives associated with camera sensor attribution to their ears during a nice coffee break, Alessandro Piva (Univ of Florence, Amped Software), Marco Fontani (FORLab) and Massimo Iuliani (Univ of Florence, Amped Software) for generously sharing the database used in~\cite{iuliani2021leak} but also giving us feedbacks on the potential use of the hidden feature, Teddy Furon (INRIA) for proposing the torture test of feeding Adobe Lightroom with a constant RAW image, Jessica Fridrich (Binghamton University) for helping us to analyze different hypotheses regarding the potential uses of a random pattern during the development process, Francis Bas for testing Adobe Lightroom on Windows, and Robert Christensen from Adobe for his very informative feedbacks, and finally the Leica shop in Lille for sharing a Q2 camera to perform new RAW and JPEG acquisitions.

This work received funding from the European Union's Horizon 2020 research and innovation program under grant agreement No 101021687 (project “UNCOVER”) and the French Defense \& Innovation Agency.

\bibliographystyle{unsrt}
\bibliography{prnu}

\end{document}